\documentclass[useAMS,usenatbib,usegraphicx]{mn2e}
\usepackage{amsmath,amsfonts,amssymb}

\newcommand{\hm}{\mbox{$h^{-1}$}}

\newcommand{\bj}{\mbox{$b_{\rm J}$}}
\def\Msun{\hbox{$\rm\, M_{\odot}$}}

\title[MCMC in Semi-Analytic Models of Galaxy Formation]{Monte Carlo
  Markov Chain Parameter Estimation in Semi-Analytic Models of Galaxy
  Formation}

\author[Henriques et al.]
{Bruno M. B. Henriques$^{1}$\thanks{E-mail: b.m.henriques@sussex.ac.uk},
 Peter A. Thomas$^{1}$ , Seb Oliver $^{1}$, Isaac Roseboom$^{1}$\\
 {}$^{1}$Astronomy Centre, University of Sussex, Falmer, Brighton BN1 9QH,
 United Kingdom}

\begin{document}

\date{Accepted 2009 March 3}

\volume{396} \pagerange{535--547} \pubyear{2009}

\maketitle

\label{firstpage}

\begin{abstract}

We present a statistical exploration of the parameter space of the
De Lucia and Blaizot version of the Munich semi-analytic (SA)
model built upon the Millennium dark matter simulation. This is
achieved by applying a Monte Carlo Markov Chain method to constrain
the 6 free parameters that define the stellar and
black-hole mass functions at redshift zero.  The model is tested
against three different observational data sets, including the galaxy
$K$-band luminosity function, $B-V$ colours, and the black hole-bulge mass
relation, separately and combined, to obtain mean values, confidence
limits and likelihood contours for the best fit model.  Using each
observational data set independently, we discuss how the SA model
parameters affect each galaxy property and find that there are strong
correlations between them.  We analyse to what extent these are simply
reflections of the observational constraints, or whether they can lead
to improved understandings of the physics of galaxy formation.

When all the observations are combined, we find reasonable agreement
between the majority of the previously published parameter values and
our confidence limits.  However, the need to suppress dwarf galaxy
formation requires the strength of the supernova feedback to be
significantly higher in our best-fit solution than in previous work.
To balance this, we require the feedback to become ineffective in
halos of lower mass than before, so as to permit the
formation of sufficient high-luminosity galaxies: unfortunately, this
leads to an excess of galaxies around L$^*$.  Although the best-fit is
formally consistent with the data, there is no region of parameter
space that reproduces the shape of galaxy luminosity function across
the whole magnitude-range.

For our best fit we present the model predictions for the \bj-band
luminosity and stellar mass functions.  We find a systematic
disagreement between the observed mass function and the predictions
from the $K$-band constraint, which we explain in light of recent
works that suggest uncertainties of up to 0.3 dex in the mass
determination from stellar population synthesis models.

We discuss modifications to the semi-analytic model that might
simultaneously improve the fit to the observed mass function and
reduce the reliance on excessive supernova feedback in small halos.
\end{abstract}


\section{Introduction}
\label{intro}

The combination of the well established $\Lambda$CDM paradigm and the
geometrical growth of computer power in recent years, has allowed
direct N-body simulations to give us a comprehensive picture of the
formation and evolution of dark matter structure, from the primordial
gravitational instabilities to the formation of massive superclusters
today \citep[The Millennium Run,][]{Springel2005}.

However this progress is not reflected in our theoretical
understanding of the behaviour of the baryons, with much of the
physics governing galaxy formation and evolution still poorly
understood.  A combined simulation of gas and dark matter resolving
small-scale galaxy evolution processes (such as gas cooling, star
formation and feedback) over a cosmologically interesting volume is
still years away.  Therefore the only plausible method to construct a
model galaxy population for comparison with observed large-scale
surveys is using a semi-analytic formalism.  Originally introduced by
\citet{White1978}, this formalism treats the dark matter structure
using either a Press-Schechter \citep{Press1974}, Monte-Carlo or
N-body approach and on top of that follows galaxy evolution using
parameterised equations governing the laws of subgrid physics.  The
basic methodology was set by \citet{Cole1991}, \citet{Lacey1991} and
\citet{White1991}, including the dependence of gas cooling and star
formation on the dark matter halo density profile, feedback and
chemical enrichment to account for the effect of supernova (SN)
explosions on the properties of the hot gas, and stellar population
synthesis models to convert star formation histories into observed
stellar properties.

Further developments included newly-derived stellar population models
and improved star formation and SN feedback laws \citep{Kauffmann1993,
Lacey1993, Cole1994}.  With the level of complexity achieved, they
were able to predict a large range of galaxy properties such as
star-formation rates, luminosity functions and relations between
circular velocity, luminosity, metallicity and mass-to-light ratios.

From this original recipe, two models started evolving separately, one
mainly based in Munich and another in Durham.  By the end of the
decade, most of the modern day prescriptions were already introduced
\citep{Kauffmann1999a, Cole2000}, including gas cooling, star
formation, chemical enrichment and dust extinction, calculations of
disks and bulge properties, stellar population synthesis models,
merger follow up with dynamical friction and an early version of the
SN feedback treatment. 

In \citet{Kauffmann2000} a model for the growth of black holes (BHs)
due to 
instabilities arising from mergers was proposed, and
\citet{Benson2003} and \citet{Delucia2004} studied new treatments of
the SN feedback, including the current model where the SN can not only
reheat the cold gas into the hot phase but also eject gas from the
halo (to be reincorporated at later times).
In parallel, a number of other groups begun to develop independent models,
to study different aspects of galaxy formation
\citep{Somerville1999,Menci2002,Hatton2003,Daigne2004,Monaco2004,Kang2005}. 

Most recently, the Munich and Durham semi-analytics 
have been combined with the Millennium
dark matter simulation and an additional recipe, the BH radio
mode, introduced to reproduce the quenching of gas cooling star
formation in the gas surrounding central cluster galaxies
(\citealt{Springel2005, Croton2006, Bower2006}; see also
\citealt{Granato2004, Cattaneo2006, Menci2006, Monaco2007, Somerville2008}).

Finally, present day studies include new dust models
\citep{Delucia2007}, the study of alternative feedback processes
such as galactic winds \citep{Bertone2007}, 
improved recipes for the stripping of gas during
galaxy mergers \citep{Font2008} and investigation of the ability of
the energy released by AGN feedback to reproduce the properties of the
intra-cluster medium \citep{Bower2008}.

With all these recipes in place, the models successfully reproduce a
vast range of observable properties, from galaxy luminosities and
colours, including environment dependences, to scaling relations such
as Tully-Fisher diagrams.  However, until now the level of agreement
with observations and the relative weight of different observations in
the final choice of the parameters has never been studied in a
statistically-consistent way.

Moreover, the large number of observational properties that the models
aim to predict requires a large number of parameters (some of which
are strongly correlated), producing considerable difficulties in
determining how to improve the agreement with new observations without
destroying the match with existing data sets. In addition, whenever reasonable
agreement proves to be impossible, it is hard to know whether there is
a failure in determining the right parameter configuration, whether
there is a fundamental problem with the underlying model, or whether
the introduction of new physics is called for.

These difficulties can be overcome by combining multiple observations
with proper sampling of high-dimensional parameter spaces.  This has
proved to be a fruitful approach in theoretical cosmology where
techniques such as Monte Carlo Markov Chain (MCMC) parameter
estimation have been extensively used (see \citet{Trotta2008}
for a comprehensive review).  The aim of this paper is to introduce
MCMC techniques into semi-analytic models of galaxy formation. While
this work was being developed, a first result was produced
by \citet{Kampakoglou2008}.  These authors have introduced similar
tools to their own semi-analytic recipe, an extension of
\citet{Daigne2004}, which uses a statistical method to generate
halos. We differ from them in that we use the semi-analytic model
\citet{Delucia2007}, in our case built upon a direct dark matter
simulation of a cosmological size (the Millennium Run). 

 This will allows us to understand how galaxy properties are
affected by individual parameters, obtain confidence limits for the
parameters, and verify the agreement between the model and different
observations in a statistically robust way.

This paper is organised as follows. In section \ref{model} we briefly
describe the semi-analytic model used in our study. In section
\ref{mcmc} we present the MCMC technique used to constrain the model
parameters and explain how it is implemented into the semi-analytic
recipe.  Section \ref{observations} describes the observational data
used in this work, clarifying which parameters are constrained by each
observational galaxy property.  In section \ref{results} we present our
results including correlations between the parameters analysed and
predictions for our best fit model. Finally in section
\ref{conclusions} we summarise our conclusions.

\section{The Model}
\label{model}

In this section we briefly describe the semi-analytic model we use for
this work \citep[hereafter DLB07]{Delucia2007}, and the underlying
dark matter simulation, the Millennium Simulation \citep{Springel2005}.

The Millennium Simulation traces the evolution of dark matter haloes
in a cubic box of 500\,\hm\,Mpc on a side. It assumes a $\Lambda$CDM
cosmology with parameters $\Omega_{\rm m}=0.25$, $\Omega_{\rm
b}=0.045$, $h=0.73$, $\Omega_\Lambda=0.75$, $n=1$, and $\sigma_8=0.9$,
where the Hubble parameter is $H_0 =
100$\,\hm\,km\,s$^{-1}$\,Mpc$^{-1}$. The simulation follows $2160^3$
dark matter particles of mass $8.6\times 10^8$\,\hm\,\Msun. Since dark
matter haloes are required to contain at least 20 particles, the
minimum halo mass is $1.7\times 10^{10}$\,\hm\Msun, with a
corresponding baryonic mass of about $3.1\times10^9$\,\hm\,\Msun.

The coupling of the semi-analytic model onto the high-resolution
N-body simulation follows the technique implemented by
\citet{Springel2001}. The treatment of physical processes driving
galaxy evolution builds on the methodology introduced by
\citet{Kauffmann1999a}, \citet{Springel2001} and \citet{Delucia2004},
and is a slightly modified version of that used in
\citet{Springel2005} and \citet{Croton2006}.

The model divides the gas content of galaxies into several distinct
phases.  When the galaxy first forms, the gas enters the {\it hot
halo} at the virial temperature.  It can then cool down to join the
{\it cold disk}.  Stars form from the disk and feedback of energy via
supernovae can cause gas to heat back up from the cold disk to the hot
halo.  Finally, it is possible to heat gas still further, ejecting it
from the galaxy into an {\it external reservoir} from which it
gradually leaks back into the hot halo.

\begin{table}
\begin{center}
\caption{Best-fit parameters for the SA model from DLB07.  The first 6
  parameters are frozen in our analysis at the values shown here.  
  A  detailed description of the parameters is given in the text.}
\label{table:sampar}
\begin{tabular}{cccccc}
\hline
$f_{\rm b}$& $z_0$& $z_{\rm r}$& $T_{\rm merger}$& $R$& $Y$\\
0.17& 8& 7& 0.3& 0.43& 0.03\\
\hline
$\alpha_{\rm SF}$ &$k_{\rm AGN}$ &$f_{BH}$ &$\epsilon_{disk}$
&$\epsilon_{halo}$ &$\gamma_{\rm ej}$\\
0.03& $7.5\times10^{-6}$& 0.03& 3.5& 0.35& 0.5\\
\hline
\end{tabular}
\end{center}
\end{table}

From the original 12 parameters in the model, we choose to freeze 6 of them at
the values chosen by DLB07, as shown in the top row of
table~\ref{table:sampar}.  The cosmic baryon fraction, $f_{b}$, is fixed
by the cosmology\footnote{Note that, as in DLB07, we use the value of
0.17 suggested by WMAP rather than the value of 0.18 used to
generate the power spectrum for the Millennium Simulation}, while the
redshifts of the beginning and end of reionization ($z_{0}$, $z_{\rm r}$)
are used to modify the baryon fraction in small halos, accounting for
the effects of photo-ionizing heating \citep{Kravtsov2004,Croton2006}.
$T_{\rm merger}$ is the threshold mass ratio that defines the
distinction between Major and Minor mergers.  $R$ is the recycled
fraction and $Y$ the yield, both of which depend upon the details of
the stellar initial mass function.  The observational data that
we use in this work do not allow us to strongly constraint any of
these values.

That leaves 6 free parameters in our study.  These are the star
formation efficiency, $\alpha_{\rm SF}$, the fraction of cold gas
accreted by the central BH during mergers, f$_{\rm BH}$, the quiescent
hot gas BH accretion rate, k$_{\rm AGN}$, the SN feedback disk
reheating efficiency, $\epsilon_{\rm disk}$, the SN feedback halo
ejection efficiency, $\epsilon_{\rm halo}$, and the ejected gas
reincorporation efficiency, $\gamma_{\rm ej}$.  We briefly describe
the meaning of each of these parameters below, for a full description
see \citet{Croton2006} and DLB07.
 
The model converts cold gas into stars at a rate given by
\begin{equation} \label{eq:sfe}
\dot{m}_{\star}
= \alpha_{\rm SF}\frac{(m_{\rm cold}-m_{\rm crit})}{t_{\rm dyn,disk}},
\end{equation}
where $m_{\rm cold}$ is the mass of cold gas, $m_{\rm crit}$ is the
mass that corresponds to a critical surface density above which
gas can collapse and form stars \citep[following][]{Kennicutt1998}, 
and $t_{\rm dyn,disk}$ is the dynamical time of the disk.  
Note that the fraction of mass locked up in stars is $(1-R)\dot{m}_{\star}dt$, 
the rest being instantaneously returned to the disk.

As massive stars complete their life cycle, SN events start injecting
energy into the surrounding medium, reheating cold disk gas and even
ejecting gas from the hot halo.  

For each mass $\Delta m_{\star}$ turned into stars, the amount of gas
reheated from the cold disk to the hot halo is given by
\begin{equation} \label{eq:edisk}
\ \Delta m_{\rm reheated}=\epsilon_{\rm disk}\Delta m_{\star}
\end{equation}
with the canonical efficiency of 3.5 being motivated by observations by
\citet{Martin1999}.

This proves insufficient to prevent star-formation in dwarf galaxies
as the cooling times are so short that the gas rapidly cools back down
to rejoin the disk.  For this reason, and motivated also by observations
of galactic outflows \citep{Martin1996}, the models allow SN to expel gas
completely from low-mass galaxies.  

The amount of energy released by supernova during the formation of
$\Delta m_{\star}$ stars is
\begin{equation} \label{eq:ehalo}
\ \Delta E_{\rm SN}=0.5\,\epsilon_{\rm halo}\,\Delta m_\star V_{\rm SN}^2
\end{equation}
where $V_{\rm SN}=630$\,km\,s$^{-1}$.  Any {\it excess}\footnote{Note that
the reheated fraction is not reduced if this excess is negative.} energy left
over from reheating the cold gas is used to eject
a mass of gas $\Delta m_{\rm ejected}$ from the galaxy
\begin{equation} \label{eq:eject}
\Delta m_{\rm ejected}=\left(\epsilon_{\rm halo} \frac{V_{\rm
    SN}^2}{V_{\rm vir}^2} -\epsilon_{\rm disk}\right) \Delta m_{\star},
\end{equation}
where $V_{\rm vir}$ is the circular velocity of the dark matter halo.

This ejected gas is kept in an external reservoir and returned to the
hot halo at a rate
\begin{equation} \label{eq:reinc}
\dot{m}_{\rm ejected}=-\gamma_{\rm ej}\frac{m_{\rm ejected}}{t_{\rm dyn}},
\end{equation}
where $m_{\rm ejected}$ is the mass of ejected gas and $t_{\rm dyn}$
is the dynamical time of the halo.

SN feedback is ineffective in large galaxies and so another
form of heating must be included.  Without it, central cluster galaxies appear
too massive and too blue, an aspect of the cooling flow
problem, well-known to X-ray astronomers.  The solution is thought to
be mechanical heating by BHs accreting at well below the
Eddington limit, with  the amount of energy released depending on the mass
accretion rate of the central supermassive BH, which in turn
depends on the BH mass.

To describe this process it is necessary to introduce two different
modes of AGN activity: the quasar and radio modes.  The former is
thought to be inefficient at heating the gas but is primarily
responsible for BH growth.  It was originally introduced into
the models simply to predict the mass of central BHs and the
corresponding SA model parameter, $f_{\rm BH}$, regulates the BH
growth by accretion associated with galaxy mergers
\begin{equation} \label{eq:fbh}
\Delta m_{\rm BH,Q}=\frac{f_{\rm BH}(m_{\rm sat}/m_{\rm
    central})\,m_{\rm cold}}{1+(280\,\mathrm{km\,s}^{-1}/V_{\rm vir})^2}.
\end{equation}

The radio mode reflects the BH growth via quiescent accretion
in a static hot halo.  It may represent either Bondi accretion
directly from the hot phase, or the accretion of small quantities of
cold gas.  It is described by the phenomenological model
\begin{equation} \label{eq:kagn}
\dot{m}_{\rm BH,R}=k_{\rm AGN}\left(\frac{m_{\rm
    BH}}{10^8\,\Msun}\right)\left(\frac{f_{\rm
    hot}}{0.1}\right)\left(\frac{V_{\rm
    vir}}{200\,\mathrm{km\,s}^{-1}}\right)^3,
\end{equation}
where $m_{\rm BH}$ is the BH mass and $f_{\rm hot}$ is the mass
fraction of hot gas in the halo.

The radio mode makes a minor contribution to the growth in mass of the
BH but is assumed to generate mechanical heating at a rate
\begin{equation} \label{eq:lagn}
L_{\rm BH}=\eta\,\dot{m}_{\rm BH,R}\,c^2,
\end{equation}
where $c$ is the speed of light and the efficiency parameter $\eta$ is
frozen at 0.1.\footnote{Note that it is the combination $\eta\,k_{\rm AGN}$
that determines the heating rate so that the value of $\eta$ is
unimportant.}  This heating is used to reduce the rate at which 
gas cools from the hot halo into the cold disk.

\section{Monte Carlo Markov Chain}
\label{mcmc}

\subsection{Bayesian Monte Carlo Markov Chain Analysis}
\label{bayes}

Monte Carlo Markov Chain (MCMC) methods are a class of algorithms for
sampling a multidimensional space with a probability proportional to
the likelihood that the model describes the observational
constraints.  The following brief description follows that in
\citet{Press2007}.

A typical application of this method is when it is possible to
calculate the probability, $P(D|x)$, of a given data set, $D$, given the
values of some model parameters, $x$.  Bayes' theorem says that, given
a prior $P(x)$, the (posterior) probability of the model (which will be
sampled by the MCMC) is $\pi(x)\propto P(D|x)\,P(x)$ with an
unknown normalising constant.  The advantages of MCMC are that the
posterior distribution and correlations for the parameters in study
can be easily recovered from the sample list and the un-normalised
probability, and that the computational power required scales only
linearly with the number of parameters.

The MCMC method uses a Markov chain to step from one point in the
sample space to the next, meaning that each point is chosen from a
distribution that depends only on the preceding point (the ergodic
property).  The transition probability $p(x_2|x_1)$ for stepping from
point $x_1$ to point $x_2$ should satisfy the detailed balance equation,
\begin{equation} \label{eq:balance}
\pi(x_1)\,p(x_2|x_1)=\pi(x_2)\,p(x_1|x_2)
\end{equation}

\subsection{Metropolis-Hastings Algorithm}
\label{metropolis}
 
There are several algorithms that can produce a chain with the
required properties, the most common being the Metropolis-Hastings
algorithm \citep{Metropolis1953,Hastings1970}.  This method
requires a proposal distribution $q(x_2|x_1)$ that can assume various
shapes, as long as the chain reaches everywhere in the region of
interest.  However, an inappropriate choice can delay significantly
the convergence of the chain.  Considering the underlying probability
distribution of our parameters we choose a log-normal proposal
distribution with a width that assures that the final acceptance of
the chain is between 10\% and 40\%.

The chain is then started in a randomly selected point in parameter
space $x_1$.  A new candidate point $x_{2c}$ is selected by drawing
from the proposal distribution, and the acceptance probability
$\alpha(x_1,x_{2c})$ calculated using the formula,
\begin{equation} \label{eq:acceptance}
\alpha(x_1,x_{2c})
=\min\left(1,\frac{\pi(x_{2c})q(x_1|x_{2c})}{\pi(x_1)q(x_{2c}|x_1)}\right)
\end{equation}
The candidate point is accepted with probability $\alpha(x_1,x_{2c})$
and $x_2$ is set equal to $x_{2c}$, or rejected and the point left
unchanged $(x_2=x_1)$.

The ratio $q(x_1|x_{2c})/q(x_{2c}|x_1)$ in
equation~(\ref{eq:acceptance}) represents the prior which we assume to
be log-normal.  

\subsection{MCMC applied to the Semi-Analytic Model}
\label{mcmcsam}

Implementing the MCMC sampling approach on the semi-analytic model
parameter space raises considerable issues related not only to the
copious amount of I/O (the original recipe reads in the full
Millennium dark matter trees), but also to the volume of calculations
required to follow the evolution of over 20 million galaxies in a
cosmological volume, through more than fifty redshift slices.  At each
MCMC step the semi-analytic model needs to be run with the proposed
set of parameters, to compute the acceptance probability by comparing
the outputted galaxy properties with the observational constraints.
The size of the calculation and the number of steps required for
convergence makes it unfeasible to perform our analysis using the full
Millennium volume.

The structure of the Millennium Simulation provides a way to
circumvent this difficulty.  The output is divided into 512 files 
which have self-contained trees, with the galaxies on each treated independently.  
We choose to perform our
analysis in a single file with a mean density and luminosity function
analogous to the full Millennium box.  This assures that the parameter
study done on it is representative of the full data set.  Only the
largest galaxies with stellar masses greater than about
10$^{11}$\,\hm\,\Msun\ are not properly sampled this way.

For our best-fit parameters, we rerun the SA model on the full
simulation: these results are presented in section~\ref{results} below.

\section{Individual Observational Constraints}
\label{observations}

\subsection{Overview}
\label{overview}

The traditional semi-analytic approach, is to adjust parameters only
considering observations at redshift zero.  Following this philosophy
we select 3 independent and local observational data sets: the
$K$-band luminosity function, the colour-stellar mass relation and the
BH-bulge mass relation, to fully constraint the 6 parameters
defining galaxy masses and formation rates of stars and AGN.

In this section we present the observations used in our analysis and
we show how each individual property constraints the different
parameters by running the MCMC sampling with one observational data
set at a time. The output is analyzed using getdist, part of
the COSMOMC software package \citep{Lewis2002}, adapted to produce 1d
and 2d maximum likelihood (profile) and MCMC marginalised (posterior)
distributions.  For the independent observational properties we use
different statistical tests to assess the likelihood of the model,
reflecting the observational uncertainty and the nature of the
relation under study.

Running our sampling technique with separate observational data sets
one at a time allows us to gain insight on which degeneracies between
the different parameters are broken by each additional observation.
We start by studying the influence of varying the parameters on the
final $K$-band luminosity function.

\subsection{The $K$-band Luminosity Function}
\label{kband}

Despite being one of the most fundamental properties of a galaxy,
stellar mass is not easily derived from observations.  To get
estimates for this quantity from the observed luminosities it is
necessary to assume mass-to-light ($M$-$L$) ratios based on stellar
population synthesis models that include still poorly-understood dust
corrections, initial mass functions (IMFs) and metallicity evolution.
On the other hand, semi-analytic models directly predict mass, but to
produce observable luminosities the same crudely-established process
must be taken in the reverse direction.

This difficulty leads us to use the $K$-band luminosity function. The
$K$-band is known as a good mass indicator as it is relatively
unaffected by dust and represents a fair sample of the stellar
population.  We combine three observational studies
\citep{Cole2001,Bell2003,Jones2006}, respectively from 2DFGRS, 2MASS
and 6DFGRS, from which we build a final luminosity function. The final data
points are given by the average of the maximum and minimum number-density
estimates in each magnitude bin, with errors $\sigma_i$ equal to half
the difference between them.

\begin{figure}
\centering
\includegraphics[width=8.4cm]{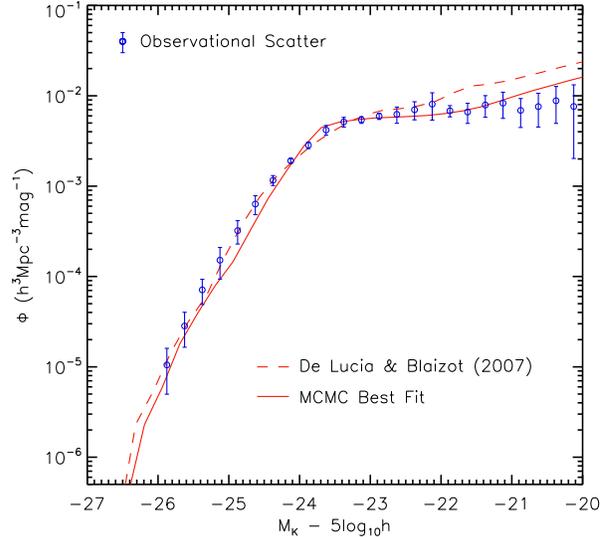}
\caption{The galaxy $K$-band luminosity function at z=0 for the DLB07
  model (dashed red line) and our best fit model (solid red line). The
  model predictions are compared with observations from
  \citep{Cole2001,Bell2003,Jones2006} combined to produce a new
  luminosity function reflecting the scatter between them.}
\label{fig:originalkband}
\end{figure}

The comparison between the $K$-band luminosity function from the
original DLB07 model (using the published parameter values) with the
observations is shown in Fig.\ref{fig:originalkband}.  The original
model already shows good agreement with the combined data except for
the faint end, over predicting the number of dwarfs galaxies with
magnitudes fainter than $K\approx -22$.

To compute the likelihood of the model for the $K$-band luminosity function
we use the chi-square probability function where
\begin{equation} \label{eq:chi}
\chi^{2}
=\sum_{i}\frac{\left(N_{i}-n_{i}\right)^2}{\sigma_i^2+n_i}
\end{equation}
is summed over the observational bin range plotted in
Fig.\ref{fig:originalkband}, and $N_i$ and $n_i$ represent
the number of observational and simulated galaxies in each bin, respectively.

\begin{figure}
\centering
\includegraphics[width=8.4cm]{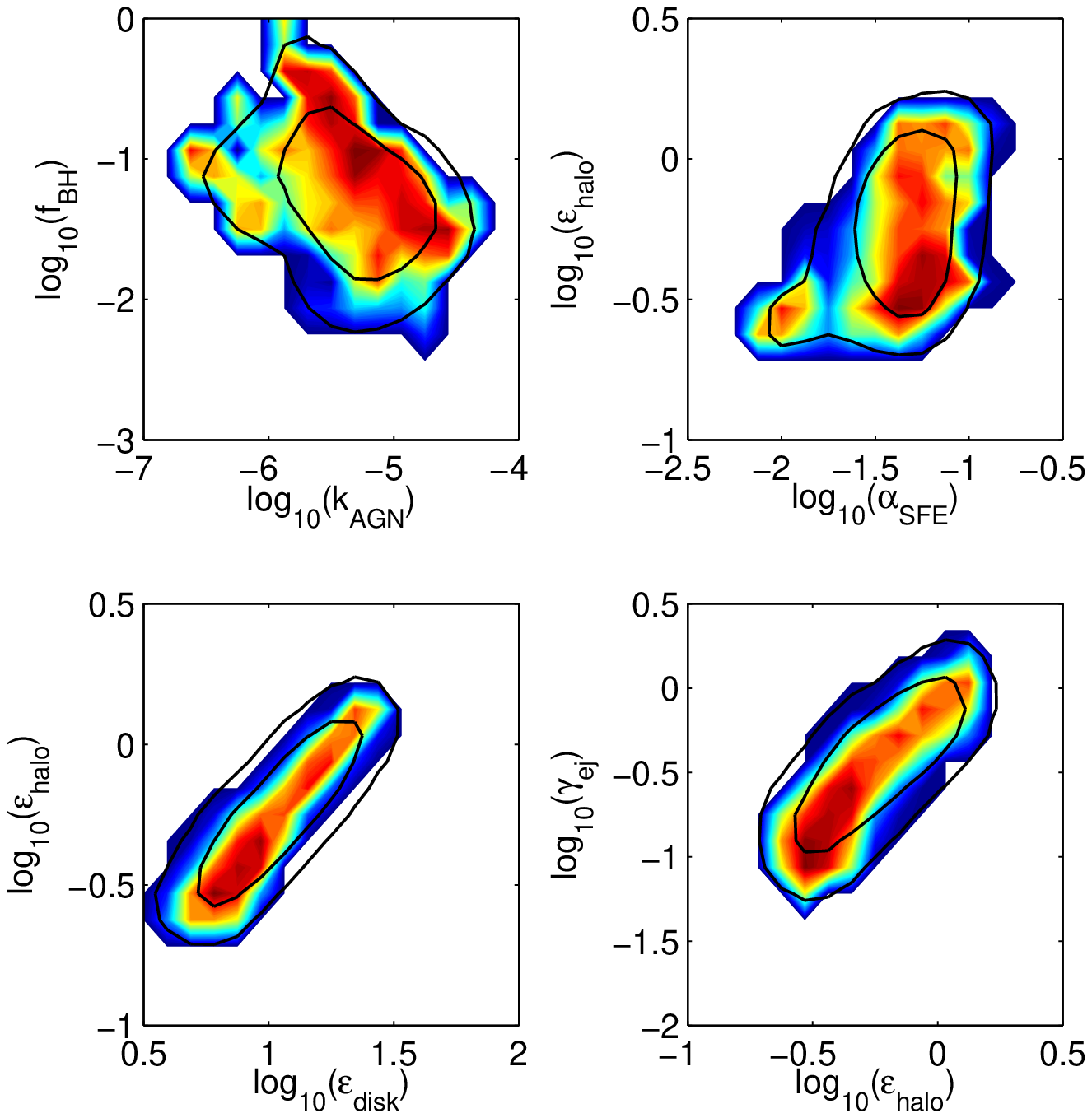}
\includegraphics[width=8.4cm]{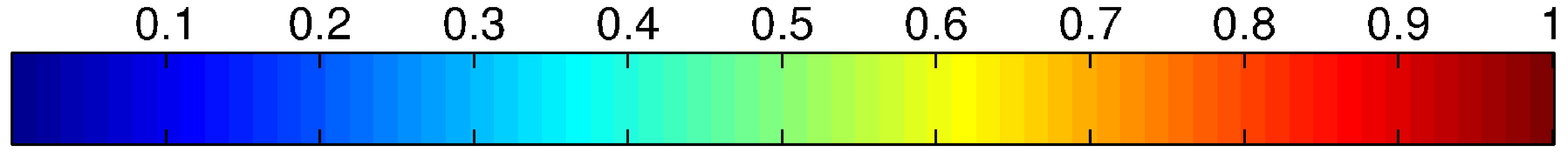}
\caption{Correlations between the 6 parameters analysed in our study
only constrained by the $\chi^2$ test on the $K$-band luminosity
function. For the values of the parameters plotted in log space, 
the solid contours represent the 68\% and 95\% preferred
regions from the MCMC (the posterior distribution)
and the colours the maximum likelihood
value sampled in each bin (the profile distribution). 
The colour scale is normalized by the maximum
likelihood value of 0.87. White regions in the plot
represent regions either with very low likelihood, less than 0.1 per
cent of the peak, or regions that have not been visited by the MCMC chain.}
\label{fig:mcmconlykband}
\end{figure}

In Fig.\ref{fig:mcmconlykband} we plot the 1$\sigma$ and 2$\sigma$
preferred values from the MCMC (solid lines) and the maximum
likelihood value sampled in each bin (colour contours), for the subset
of the original parameters (with values plotted in log space)
constrained only by the observational $K$-band luminosity function.

In interpreting this and future plots, one should bear in mind that
the contours follow the MCMC sampling in parameter space, which
should trace out the relative likelihoods of different regions 
(the posterior distribution). The colours represent 
the maximum likelihood projected along all the
hidden dimensions in the plot (profile distribution).  
Usually, as in this case, the two match
fairly well.  The exceptions arise when there is a high-likelihood
region that only occupies a small volume of parameter space.

The two lower panels of Fig.\ref{fig:mcmconlykband} show that the
three parameters controlling the SN feedback are all positively
correlated with each other, in a way that allows us to learn how the
$K$-band luminosity function constrains the feedback model.  
To do this we rewrite
equation~\ref{eq:eject} as
\begin{equation} \label{eq:ejectratio}
{\Delta m_{\rm ejected}\over\Delta m_\star}
=\epsilon_{\rm halo}\,\frac{V_{\rm SN}^2}{V_{\rm vir}^2}\,\left( 
1-{\epsilon_{\rm disk}\over\epsilon_{\rm halo}}
\,\frac{V_{\rm vir}^2}{V_{\rm SN}^2}\right),
\end{equation}
from which we see that the amount of ejected gas per unit mass of star
formation drops to zero for halos with virial speed greater than
\begin{equation} \label{eq:ejectvvir}
V_{\rm vir,0}=\left(\epsilon_{\rm halo}\over\epsilon_{\rm
  disk}\right)^{1\over2}\,V_{\rm SN}.
\end{equation}
In our analysis this cutoff is represented by the line of maximum
likelihood in the lower left panel of Figure~\ref{fig:mcmconlykband}
and corresponds to $v_{\rm vir}\approx 140$\,km\,s$^{-1}$, which
translates into $M_{\rm star}\approx 10^{10.5}M_{\odot}$ and
$M_K\approx -23$.  This cutoff virial velocity is lower than in DLB07,
which means that our SN feedback stops being effective at fainter
magnitudes, allowing more stars to form in $L_{\star}$ galaxies.  Since
we need to assume a stronger SN feedback to decrease the faint end of
the luminosity function, this is the only way to assure that enough
stars will still form in brighter galaxies.

For a given value of $V_{\rm vir,0}$ the amount of ejected gas is
proportional to $\epsilon_{\rm halo}$, with the maximum likelihood
solutions showing a linear relation $\gamma_{\rm ej}\propto
\epsilon_{\rm halo}$.  This corresponds to a roughly constant amount
of gas being held in the external reservoir: in a steady state the
external gas content is proportional to the ratio of the influx and
outflux rates.  Our regions of high likelihood represent a
considerably higher amount of gas being in the external reservoir than
in DLB07 with a corresponding reduction in star formation in faint
galaxies.

The value of $\epsilon_{\rm disk}$, controlling the reheating of cold
to hot gas, has a minor impact except as a way of controlling the
critical magnitude limit above which feedback is ineffective.
Presumably cooling times are so short in galaxies below the magnitude
limit that any gas that is reheated will quickly cool down again.  One
might have expected that it could be used to control the stellar mass
fraction in large galaxies where the cooling time is relatively long,
however this does not seem to be the case.

The AGN feedback parameters, shown in the top-left panel of
Figure~\ref{fig:mcmconlykband}, have a broader acceptable region but
also show a high-likelihood spine that runs diagonally down from
top-left to bottom right.  This can be explained by combining
equations~\ref{eq:fbh}, \ref{eq:kagn} and \ref{eq:lagn} to obtain the
mechanical heating rate produced by this process,
\begin{equation}\label{eq:agnheat}
L_{\rm BH}\propto f_{\rm BH}\,k_{\rm AGN}\,m_{\rm cold}\,f_{\rm hot}.
\end{equation}
Thus, for given cold and hot gas fractions in the galaxies, the line
represents a single heating rate. This degeneracy is broken
if the BH masses are used as a constraint (section~\ref{bhbm}), 
since their growth is mainly dominated by the quasar mode.

The maximum likelihood channels described above all have star-formation
efficiencies of $\epsilon_{\rm SF}\approx0.04$, similar to that of DLB07.
This reinforces the conclusion that the SN and AGN feedback parameters act
to maintain a constant mass of cold gas available for star-formation.

Apart from the main band discussed above, the two upper panels in
Figure~\ref{fig:mcmconlykband} show alternative regions of
acceptable likelihood. These have lower star-formation efficiencies,
requiring a greater mass of cold gas.  This, in turn, leads to a
smaller product of $f_{\rm BH}$ and $k_{\rm AGN}$.  We do not dwell on
these solutions here as they seem to be ruled out by other
observations: in particular they result in excessive BH-bulge
mass ratios.

\subsection{The Colour-Stellar Mass Relation}
\label{cm}

The star formation rate is another essential quantity in
characterising the galaxy population. Although this property can be
directly extracted from the models, it is not easily comparable with
observations.  While in the models, the mass transformed into stars at
each time step is computed, only indirect observational estimators
are available.

Galaxy colours are one such indirect measure of the recent
star-formation history of galaxies, with a clear bimodality between an
old, passively evolving red population, and a young star forming blue
sequence \citep[e.g.][]{Kauffmann2003, Brinchmann2004, Baldry2004}.

\begin{figure}
\centering
\includegraphics[width=8.4cm]{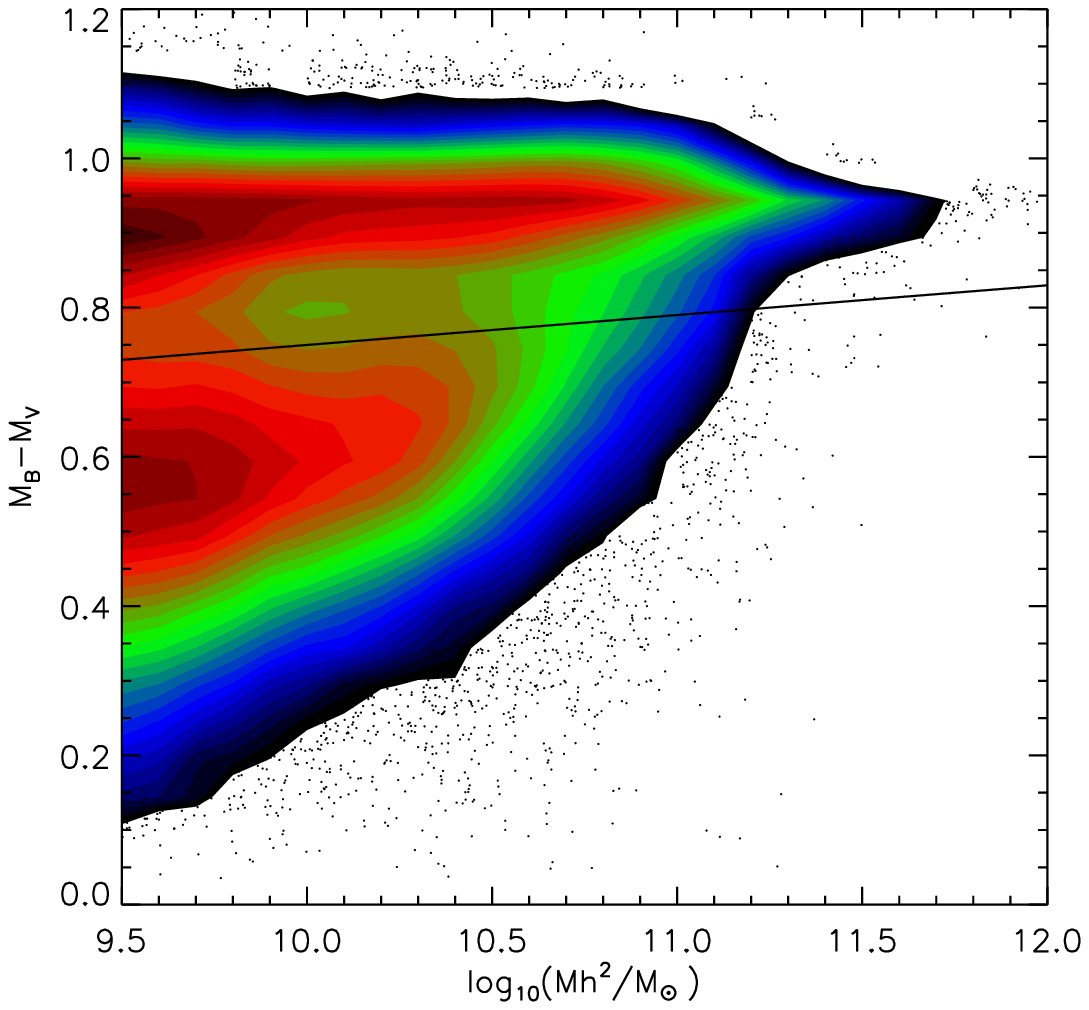}
\includegraphics[width=8.4cm]{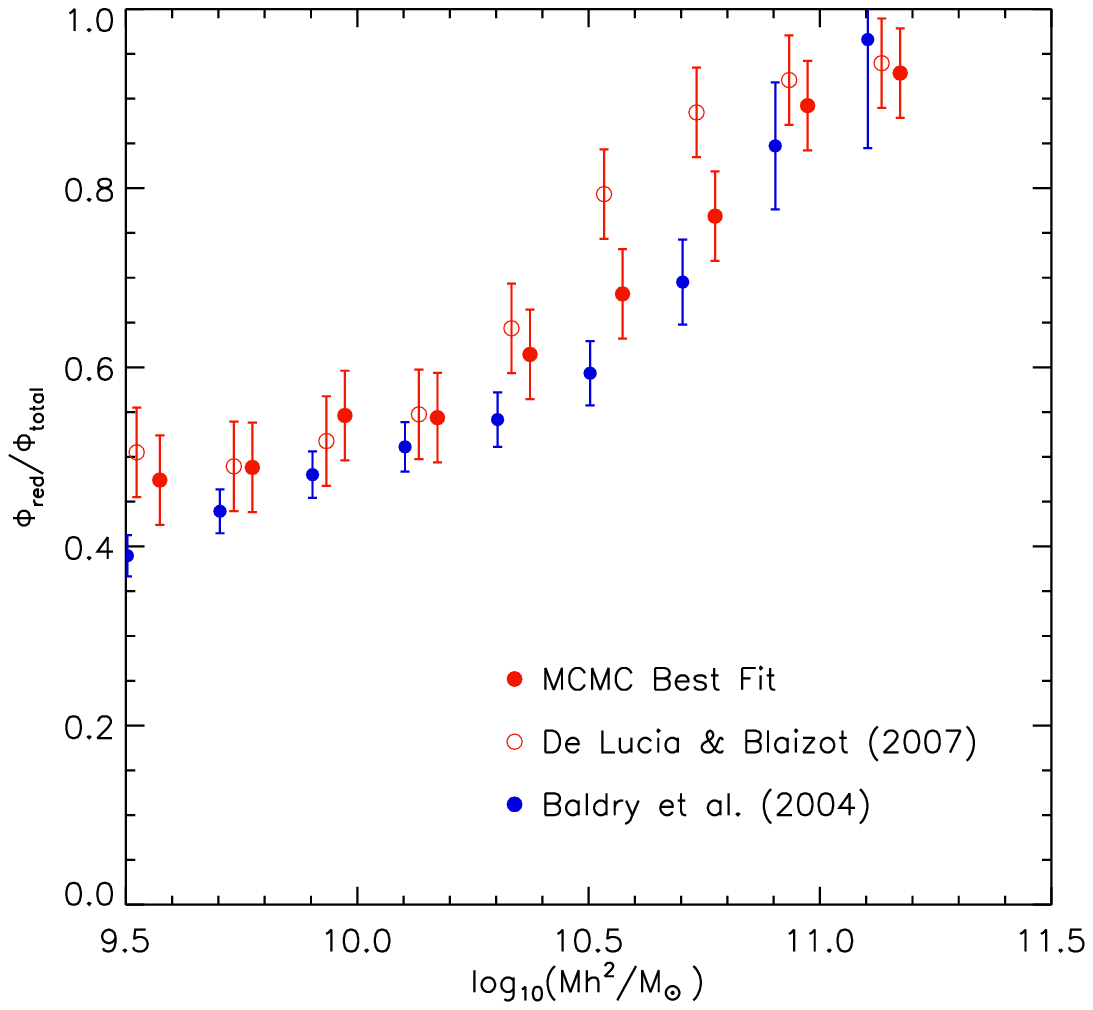}
\caption{Comparison between model and observational colours. The upper
panel shows the DLB07 B-V colour-stellar mass relation, with the solid
line representing the empirical division between populations in the
model.  On the bottom panel the fraction of red galaxies as a function
of stellar mass from DLB07 (red open circles) and our best fit model
(red filled circles) is compared with
observations from \citet{Baldry2004} (filled blue squares).}
\label{fig:colorfract}
\end{figure}

The top  panel of Fig. \ref{fig:colorfract} shows the colour-magnitude 
relation  for the DLB07 model.  Although it has  some problems in
correctly predicting the slope and fraction of each population in some
mass  ranges  \citep[see][]{Baldry2006,Weinmann2006a}, it clearly
reproduces the  bimodality.

To test the correctness of model colours, we divide the galaxies into
the two populations using the selection criteria in
\citet{Weinmann2006a}, $(g-r)=0.7-0.032\,(M_r-5\log h+16.5)$, converted
into a cut on the colour-stellar mass relation at redshift zero,
$(B-V)=0.065\,\log({\rm M}_{\star}h^2/\Msun)+0.09$, and shown as
the solid line in the upper panel of Fig. \ref{fig:colorfract}.
The conversion from the $g-r$ to the $B-V$ colour was done
  following \citet{Fukugita1996}, $g-r=1.05(B-V)-0.23$.  The fraction
of red galaxies for different mass bins is then compared with
observations from \citet{Baldry2004} as shown in the lower panel.  The
observational masses based on the 'diet' Salpeter IMF \citep{Bell2003}
were reduced by 0.15 dex to agree with the IMF assumed in our
semi-analytic model \citep{Chabrier2003}.  The fact that a
  different blue band was used in the observational colour cut ($u-r$
  instead of our converted $g-r$) could potentially lead to
  discrepancies in the number of objects identified in each population
  if the two gaussian distributions defining each population
  significantly overlap.  However, as Fig. \ref{fig:colorfract} shows,
  there is a clear division between red and blue galaxies in the $B-V$
  CM diagram for model galaxies.  In this way any small differences in
  the number of red galaxies caused by the different colour cut used,
  should be well within the 0.05 error assumed for the model
  fraction.

Following \citet{Croton2006} we take the resolution limit for model
colours to be at a stellar mass of approximately
$10^{9.5}\,h^{-2}\,M_{\odot}$, above which the model reproduces the
red fraction reasonably well, with some minor excess for galaxies
above $L_{\star}$.

The agreement between model and observational colours is calculated
using a maximum likelihood method with a constant value for the errors
in the model ($\sigma_{model}=0.05$) given by the variation in the
fraction of red galaxies in a sample of 20 sub-volumes of the
Millennium Simulation, similar to the one used in our analysis.

We assume that both model and observational values are Gaussian
distributed around the true fraction $F$, with a likelihood
\begin{equation} \label{eq:biglikecolor}
\mathcal{L}_{(\rm Colour)} = \exp\left\{
-\frac{\left(f_{\rm model}-F\right)^2}{2\sigma_{\rm model}^2}
-\frac{\left(f_{\rm obs}-F\right)^2}{2\sigma_{\rm obs}^2} \right\}
\end{equation}
that has a maximum value
\begin{equation} \label{eq:likecolor}
\mathcal{L}_{(\rm Colour)} = \exp\left\{
-\frac{\left(f_{\rm model}-f_{\rm obs}\right)^2}
{2(\sigma_{\rm model}^2+\sigma_{\rm obs}^2)}\right\}. 
\end{equation}

\begin{figure}
\centering 
\includegraphics[width=8.4cm]{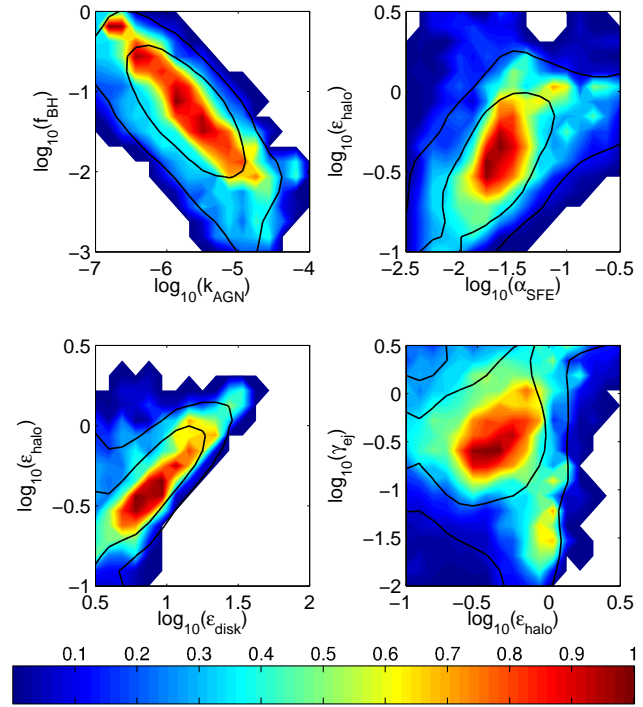}
\includegraphics[width=8.4cm]{colourbar.ps}
\caption{As for Figure~\ref{fig:mcmconlykband}, but constrained only
  by the maximum likelihood test on the fraction of red galaxies.  The
  colour scale is normalized by the maximum likelihood value of 0.89.}
\label{fig:mcmconlycolour}
\end{figure}

The DLB07 model, tuned to reproduce a different set of observational
colours (the red and blue luminosity functions from 2DFGRS), correctly
predicts the fraction of red galaxies at low and high masses but makes
the transition from moderate to high red fractions at a stellar
mass which is too low.

In Fig~\ref{fig:mcmconlycolour} we plot the allowed regions in
likelihood and posterior space. 
Perhaps surprisingly, the colour constraint picks
out a similar relationship between $\epsilon_{\rm halo}$ and
$\epsilon_{\rm disk}$ as does the $K$-band.  This is because it also
requires a cessation of SN heating in galaxies with virial speeds
above 140\,km\,s$^{-1}$ which would otherwise have an excessive red fraction.

The constraint again requires a constant mechanical heating from AGN
feedback (as shown by the line of maximum likelihood in the upper-left
panel of the figure) which is responsible for the elimination of blue
galaxies at high masses.  The line of highest likelihood lies slightly
below that seen for the $K$-band constraint.  Along with this, the
upper-right panel of the figure shows a preference for a slightly
lower star-formation efficiency.  However, in each case there is an
acceptable region where the allowed parameter spaces overlap.  This changes
when we move to our third constraint.

\subsection{The Black Hole-Bulge Mass Relation}
\label{bhbm}

We have seen in the previous section that the power of the radio-mode
AGN feedback depends upon the product of the quasar and radio-mode
growth factors.  However, the mass growth of the BHs is
dominated by the quasar mode alone.  We can therefore use the
BH-bulge mass relation to break this degeneracy.

If we require semi-analytic galaxies to be constrained solely by
this relation, then the other parameters in the model will be free to
shift into implausible values allowing any point in parameter space
to have a reasonable likelihood.  For this reason, we require model
galaxies to follow both the BH-bulge mass relation and the
$K$-band luminosity function of bright galaxies (i.e.~the host
galaxies of these BHs).

\begin{figure}
\centering
\includegraphics[width=8.4cm]{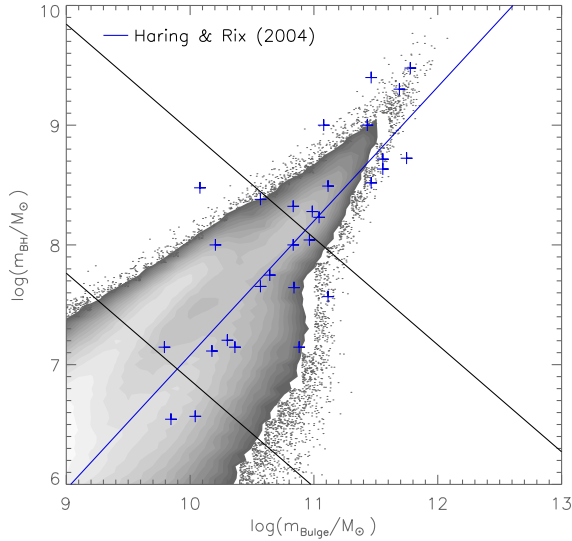}
\caption{The BH-bulge mass relation for DLB07 (contours and
  dots) is compared with local observational data from
  \citet{Haring2004} (blue crosses).  The best fit to the
  observational data points is given by the blue line running from
  bottom-left to top-right, while the two black lines perpendicular to
  this relation divide galaxies into the two mass bins used to compute the
  likelihood.}
\label{fig:originalbbh}
\end{figure}

The BH and bulge masses for the original model are plotted in
Fig. \ref{fig:originalbbh} and compared with local observations from
\citet{Haring2004}. The model galaxies fall on top of the
observational best fit (given by the blue line), with the scatter in
the relation also reproduced.

In order to test the semi-analytic results against observations, we
divide the data into two bins (perpendicular to the
observational best fit), represented by the solid black lines on
Fig.~\ref{fig:originalbbh}, $15.2<m_{\rm BH}+0.90\,m_{\rm
  Bulge}\leq17.75$ and $17.75<m_{\rm BH}+0.90\,m_{\rm Bulge}$, and for
each of the bins we compute the binomial probability for the observed 
distribution of mass ratios above and below the best-fit line, given
the fractional distribution from the model galaxies:
\begin{equation} \label{eq:likebhb}
\mathcal{L}_{(\mathrm{BH-Bulge})}=\left\{\begin{array}{ll}
2I_p(k,n-k+1),& I_p\leq0.5\\
2(1-I_p(k,n-k+1)),& I_p>0.5\end{array}\right.
\end{equation}
where $k$ is the number of observed galaxies above the best fit in
each bin, $n$ the total number of observed galaxies in the same bin,
and $p$ is the equivalent fraction, $k/n$, for the model galaxies in
the bin. $I_p(a,b)$ is the incomplete beta function as defined in
\citet{Press2007}. The two formulae are required since
we need to exclude both extremes of the distribution, corresponding
to an excess of points both above and bellow the best-fit line.

\begin{figure}
\centering
\includegraphics[width=8.4cm]{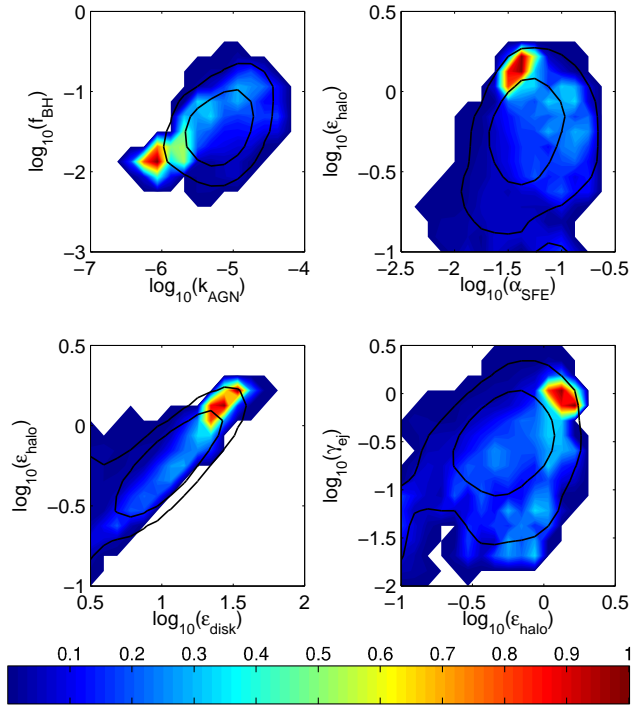}
\includegraphics[width=8.4cm]{colourbar.ps}
\caption{As for Figure~\ref{fig:mcmconlykband}, but constrained only
  by the binomial test on the BH-bulge mass relation and by
  the $K$-band luminosity function of galaxies above $M_k=-23$.  The
  colour scale is normalized by the maximum likelihood value of 0.86.}
\label{fig:mcmconlybhbm}
\end{figure}

In Fig. \ref{fig:mcmconlybhbm} we plot the posterior and 
profile likelihood distributions
of the parameters constrained only by this binomial test and the
$K$-band luminosity function of galaxies brighter than $M_k=-23$. 

It is immediately apparent that the region of high likelihood is much
smaller for this test than for the first two constraints.  Moreover,
this region corresponds to high values for the SN feedback parameters,
combined with a low AGN feedback efficiency.  As such, it is
incompatible with the acceptable regions in those previous tests.

However, there is a lower-likelihood, but still acceptable (likelihood
$> 0.1$) region that extends towards lower SN and higher AGN
parameters.  As the MCMC contours show, this occupies a much larger
volume of parameter space than the high-likelihood peak (and so, in a
Bayesian sense, the true solution is more likely to be found in the
former than the latter).

Looking at the upper-left panel in the figure, we can see that the
acceptable region runs from bottom-left to top-right, perpendicular to
the lines seen in the previous two tests.  The BH-bulge mass ratio
thus breaks the degeneracy in the AGN parameters.

\section{Combined observational constraints}
\label{results}

\begin{figure}
\centering
\includegraphics[width=8.4cm]{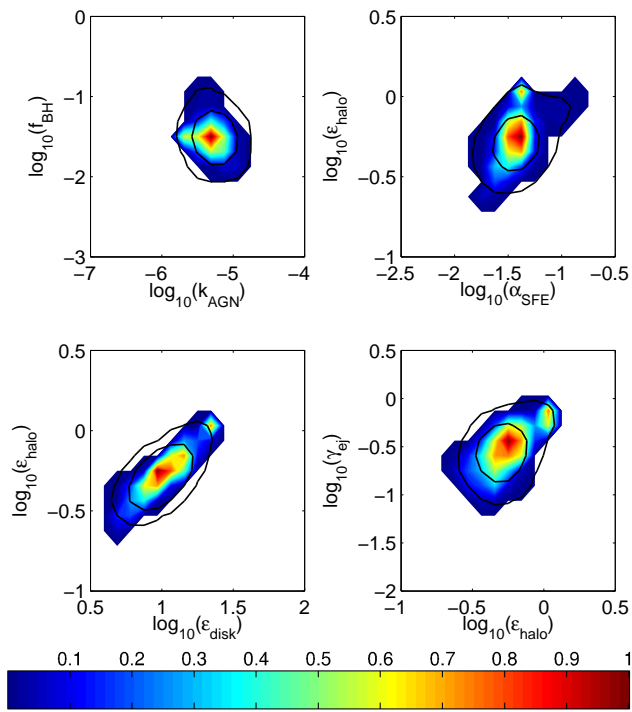}
\includegraphics[width=8.4cm]{colourbar.ps}
\caption{As for Figure~\ref{fig:mcmconlykband}, but constrained by all
  three observational properties: the $K$-band luminosity function,
  the fraction of red galaxies, and the BH-bulge mass relation.  The
  colour scale is normalized by the maximum likelihood value of
  0.037.}
\label{fig:mcmc2d}
\end{figure}

\begin{figure}
\centering
\includegraphics[width=8.4cm]{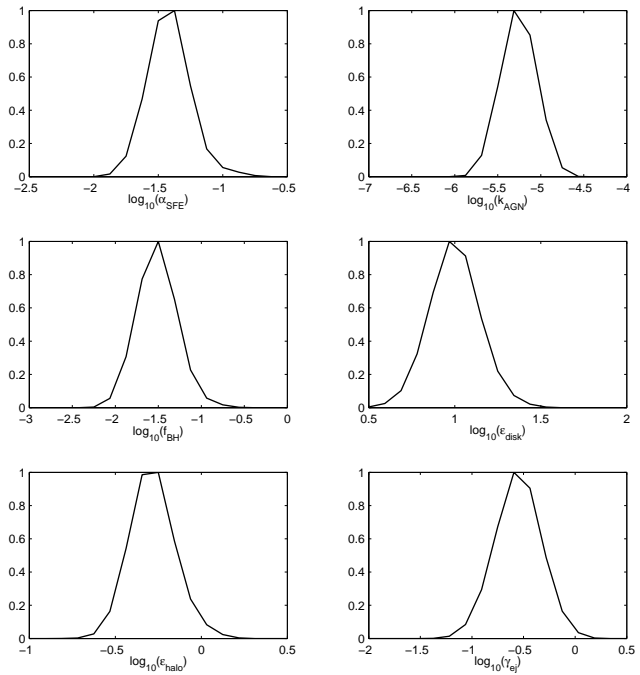}
\caption{Likelihood distributions for the 6 parameters studied.  The
solid lines represent the maximum likelihood in each bin marginalised
over the other dimensions in parameter space.}
\label{fig:mcmc1d}
\end{figure}

The likelihood of the model for a given point in parameter space is
computed by taking the product of the three independent observational
constraints described in the previous sections.
\begin{equation} \label{eq:liketotal}
\pi(x_i)=\mathcal{L}_{(\rm K-Band)}\times\mathcal{L}_{(\rm Colour)}
\times \mathcal{L}_{(\mathrm{BH-Bulge})}
\end{equation}  

This quantity is calculated at each MCMC step and used to derive the
acceptance probability (equation~\ref{eq:acceptance}).  We run our
sampling over approximately 30\,000 steps, excluding an initial
burn-in to assure the independence of the final results from the
starting point of the chain.  The output is then analysed using
getdist, part of the COSMOMC software package \citep{Lewis2002}.
As for the individual constraints, the code is adapted to
produce 1d and 2d maximum likelihood (profile) and MCMC marginalised
(posterior) distributions, best values and confidence limits for the
parameters, convergence statistics for the chain and correlation
statistics for each parameter.

Although, in our sampling we do not impose rigid limits on the
parameter range, Figure~\ref{fig:mcmc2d} shows that the preferred
regions are well-constrained.  Figure~\ref{fig:mcmc1d} shows the
profile distribution for each parameter, marginalised over
all others.  Unfortunately, the maximum likelihood is just 0.037, thus
the best-fit solution is incompatible with the three combined
observations at the 2-$\sigma$ level.

The principal cause of the low likelihood is an incompatibility
between the BH-bulge mass constraint and the $K$-band and $B-V$ colours.  
As discussed in section \ref{conclusions}, this might be caused by 
observational uncertainties in the black hole and bulge masses or
might reflect a deficiency in the black hole growth model, which 
is still very simplistic. 
Nevertheless, observational uncertainties, principally that associated with stellar
population synthesis modelling, make it premature to conclude that
the DLB07 formalism is ruled out.

\subsection{Best Fit Parameters and Confidence Limits}
\label{bestfitparams}

\begin{table*}
\begin{center}
\caption{Statistics from the MCMC parameter estimation for the 6
parameters selected from the original model. The best fit and
confidence limits (derived from the colour contours in Fig. \ref{fig:mcmc2d}) are compared with the published values from DLB07.}
\label{table:margestats}
\begin{tabular}{lccccccc}
\hline
\hline
&DLB07 &Mean  &lower (2$\sigma$ limit)  &lower (1$\sigma$ limit)  &upper (1$\sigma$ limit) &upper (2$\sigma$ limit)\\
\hline
\hline
$\alpha_{SF}$ (SFE)&0.03 &0.039 &0.020 &0.020 &0.11 &0.13\\
\hline
$k_{AGN}$ (AGN Radio) &$7.5\times10^{-6}$ &$5.0\times10^{-6}$ &$2.4\times10^{-6}$  &$2.4\times10^{-6}$  &$9.7\times10^{-6}$ &$1.1 \times10^{-5}$\\
$f_{BH}$ (AGN Quasar)&0.03 &0.032 &0.014 &0.014 &0.103 &0.115 \\ 
\hline
$\epsilon_{disk}$ (SN Reheating)&3.5 &10.28 &4.43 &4.52 &24.37 &24.37 \\
$\epsilon_{halo}$ (SN Ejection)&0.35 &0.53 &0.26 &0.26 &1.17  &1.17\\
$\gamma_{ej}$ (SN Reincorporation)&0.5 &0.42 &0.08 &0.08 &0.73 &0.79  \\
\hline
\hline
\end{tabular}
\end{center}
\end{table*}

The best fit and confidence limits for the 6 free parameters, together
with the published values from DLB07 are shown in table
\ref{table:margestats}. All the parameter values in the original model
fall within our 2$\sigma$ regions except for the SN disk reheating
efficiency, which we require to be larger than before.

Both the star formation efficiency and the AGN quasar mode parameters
from DLB07 closely match our best fit values, while our AGN radio mode
efficiency is slightly lower than before.

For the SN feedback parameters, the original halo ejection efficiency,
$\epsilon_{\rm halo}$ is below our best fit, whereas the original gas
reincorporation efficiency, $\gamma_{\rm ej}$, is slightly higher.
This combination acts so as to produce a higher fraction of gas
trapped in the external reservoir with the new parameters, and hence a
smaller mass of cold gas available for star-formation in dwarf galaxies.

The DLB07 SN reheating efficiency $\epsilon_{disk}$, is considerably
lower than our best-fit.  As discussed earlier, the main effect of
this is to raise the ratio $\epsilon_{\rm disk}/\epsilon_{\rm halo}$
with the new parameters and hence, from equation~\ref{eq:ejectvvir},
to lower the critical virial speed above which feedback is
ineffective.

\subsection{Galaxy Properties in our Best Fit Model}
\label{galaxyproperties}

The MCMC parameter estimation was carried out using only one data-file
representing 1/512 of the Millennium volume.  In this section we
present results using our best-fit parameters in the full volume.

\subsubsection{Galaxy Luminosity Functions}
\label{bestfitkband}

\begin{figure*}
\centering
\includegraphics[width=8.4cm]{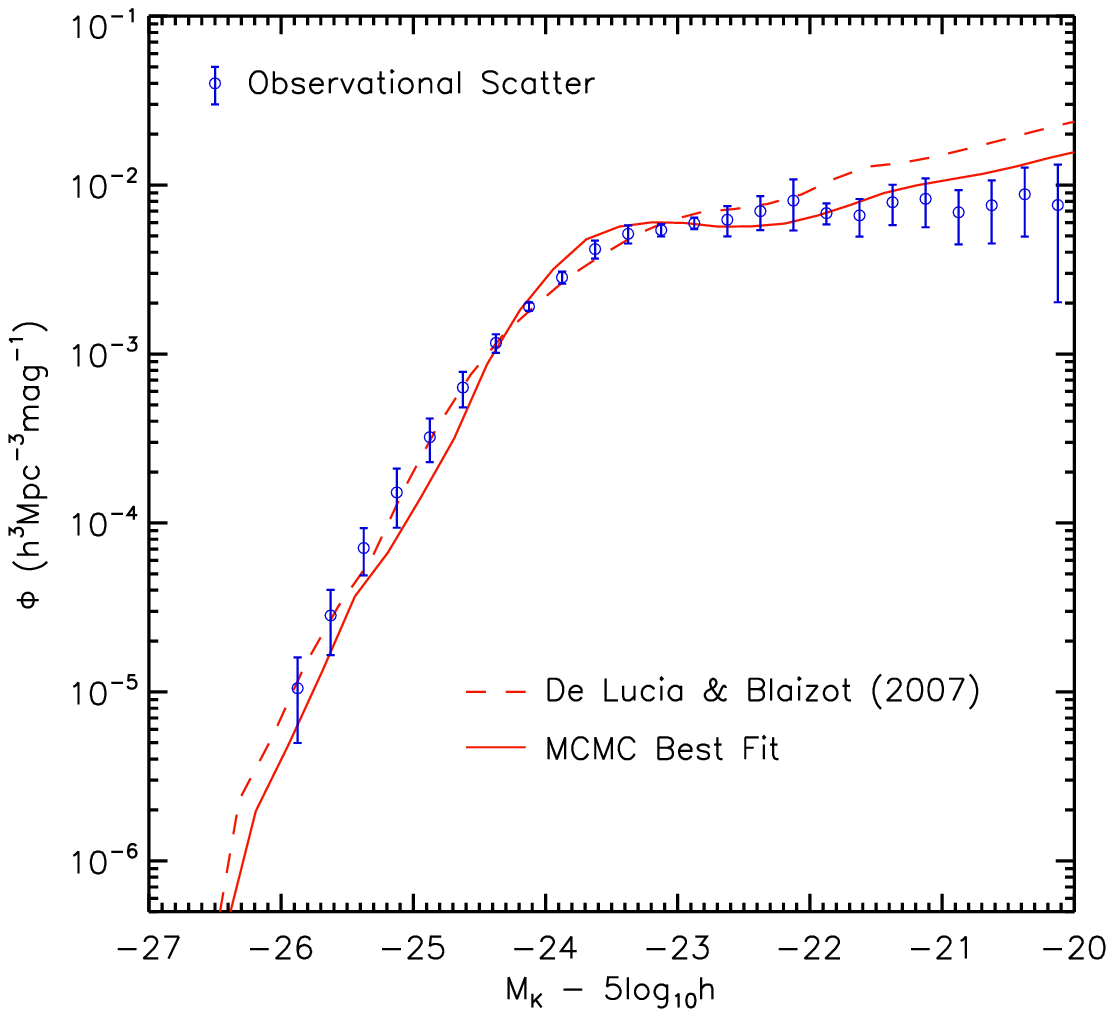}
\includegraphics[width=8.4cm]{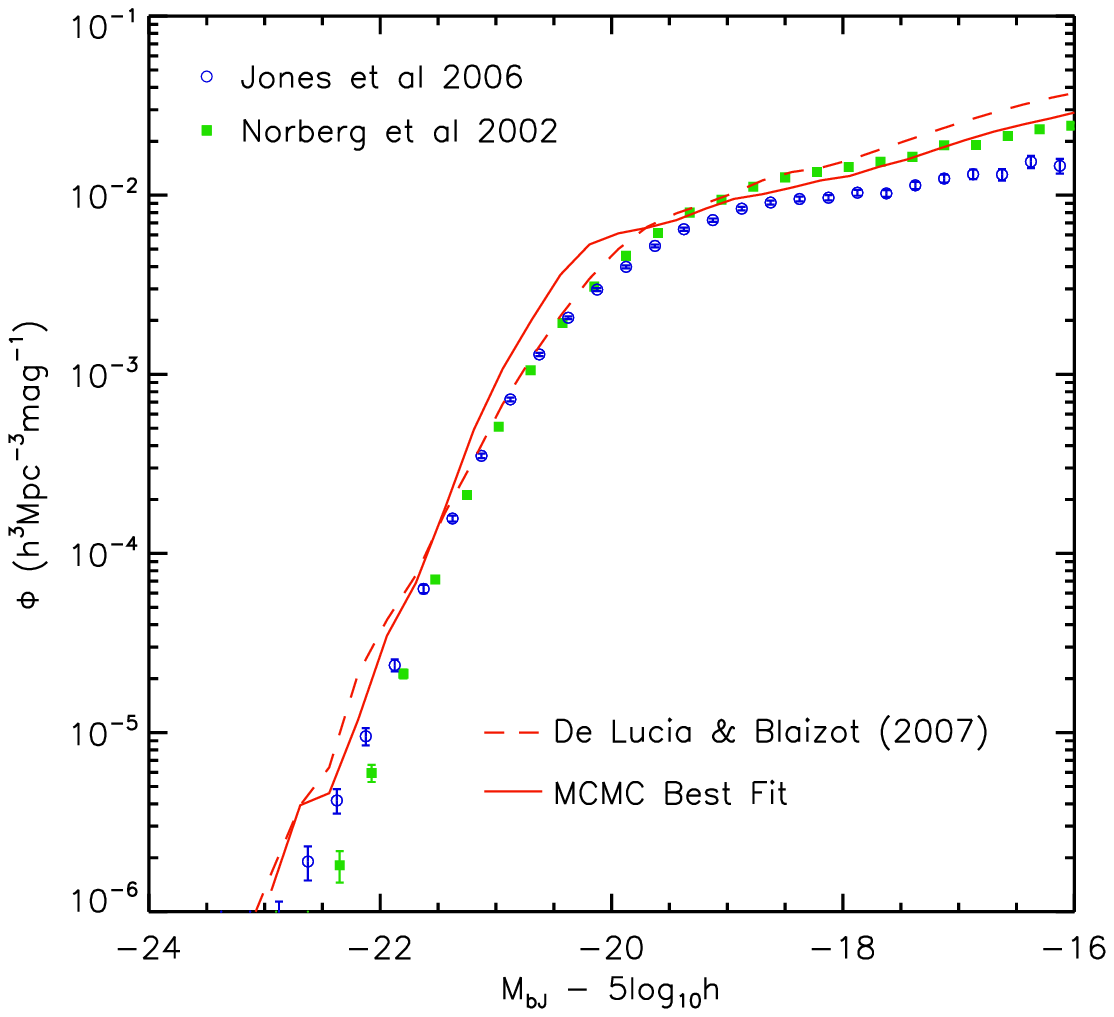}
\caption{Comparison of the predicted $K$-band (left panel) and
  \bj-band (right panel) luminosity  functions at $z=0$ from DLB07
  (dashed red line) and our best fit model (solid red line).  On the
  left panel, the data points represent the observations used to
  constrain the luminosities of galaxies in our MCMC parameters
  estimation \citep{Cole2001, Bell2003, Jones2006}.
  On the right panel, the \bj-band luminosity function
  is compared with observations from 2DFGRS (green filled squares) and
  6DFGS (blue open circles), respectively \citet{Norberg2002} and
  \citet{Jones2006}.}
\label{fig:bestfitkband}
\end{figure*}

The requirement that star formation be inefficient in low-mass
galaxies is a common problem to both sets of semi-analytic models
built upon the Millennium Simulation. There is an apparent excess of
dwarf galaxies that can be seen in the $K$-band luminosity function in
both \citet{Croton2006} and \citet{Bower2006}.

In the left panel of Fig. \ref{fig:bestfitkband} the $K$-band
luminosity functions from DLB07 and our best-fit model are plotted
against the observational data set used to constrain the sampling. 
As discussed in section \ref{kband}, to
get a good agreement with observations, the model needs to form
considerably fewer stars in low mass galaxies.  This is achieved by
reducing the amount of cold gas available for star-formation by
increasing the SN heating efficiency and decreasing the amount of
gas reincorporated at each time step. To assure that enough 
$L_{\star}$ galaxies are produced, the virial velocity cutoff above which SN
feedback is ineffective is lowered, by raising $\epsilon_{\rm
  disk}$ relative to $\epsilon_{\rm halo}$ (equation
\ref{eq:ejectvvir}).

In the right panel of Fig.\ref{fig:bestfitkband}, the best-fit model
seems to show poorer agreement with the \bj-band than the
original DLB07.  The new fit does reproduce the
number-density of dwarf galaxies accurately, but shows a large excess
around $L_{\star}$.  This is partly a reflection of the excess seen in
the $K$-band in the same region, but has a large magnitude.  Given the
good match to the colour fraction, this seems surprising and points to
inconsistencies and/or uncertainties in the conversion of mass to
luminosity via stellar population synthesis (see section \ref{bestfitsmf}).

\subsubsection{Galaxy Colours}
\label{bestfitcolours}

\begin{figure}
\centering
\includegraphics[width=8.4cm]{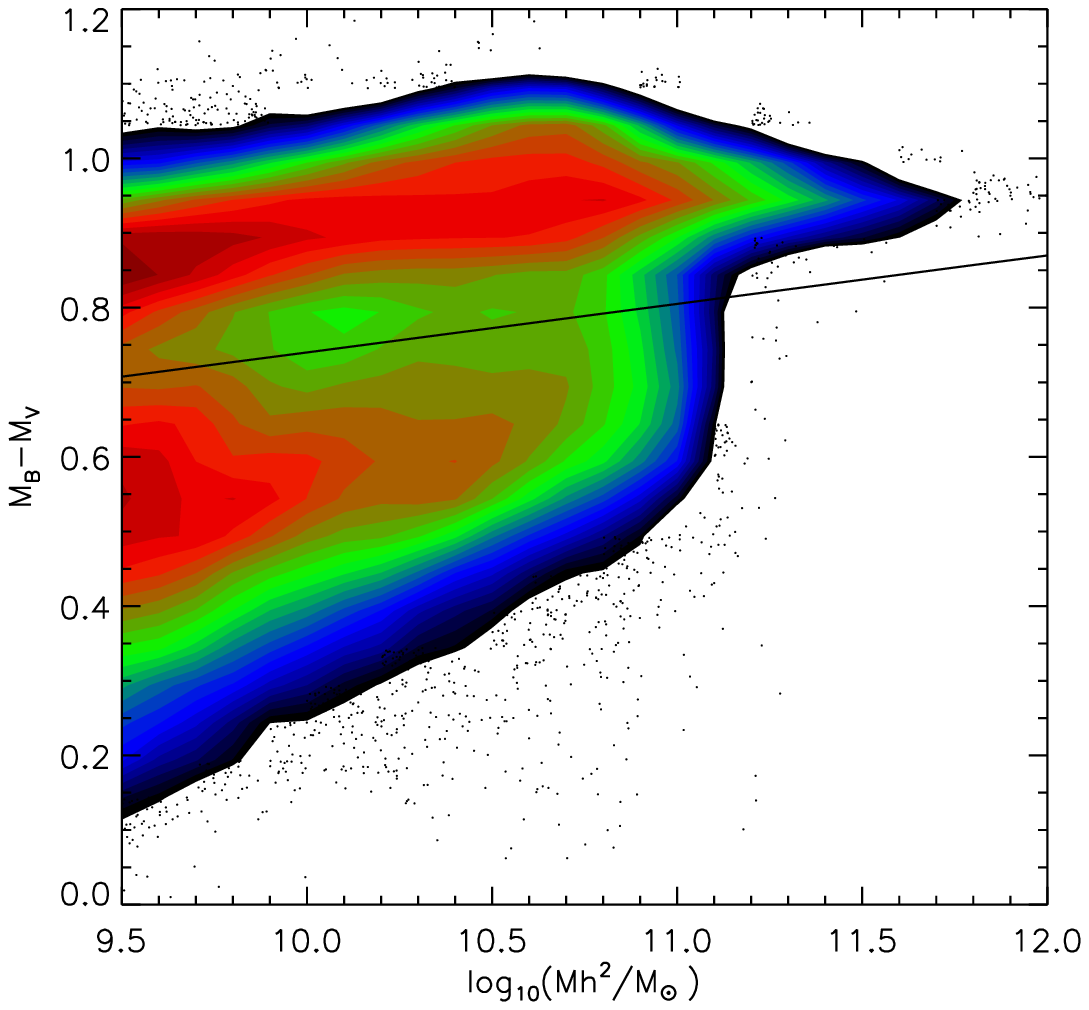}
\includegraphics[width=8.4cm]{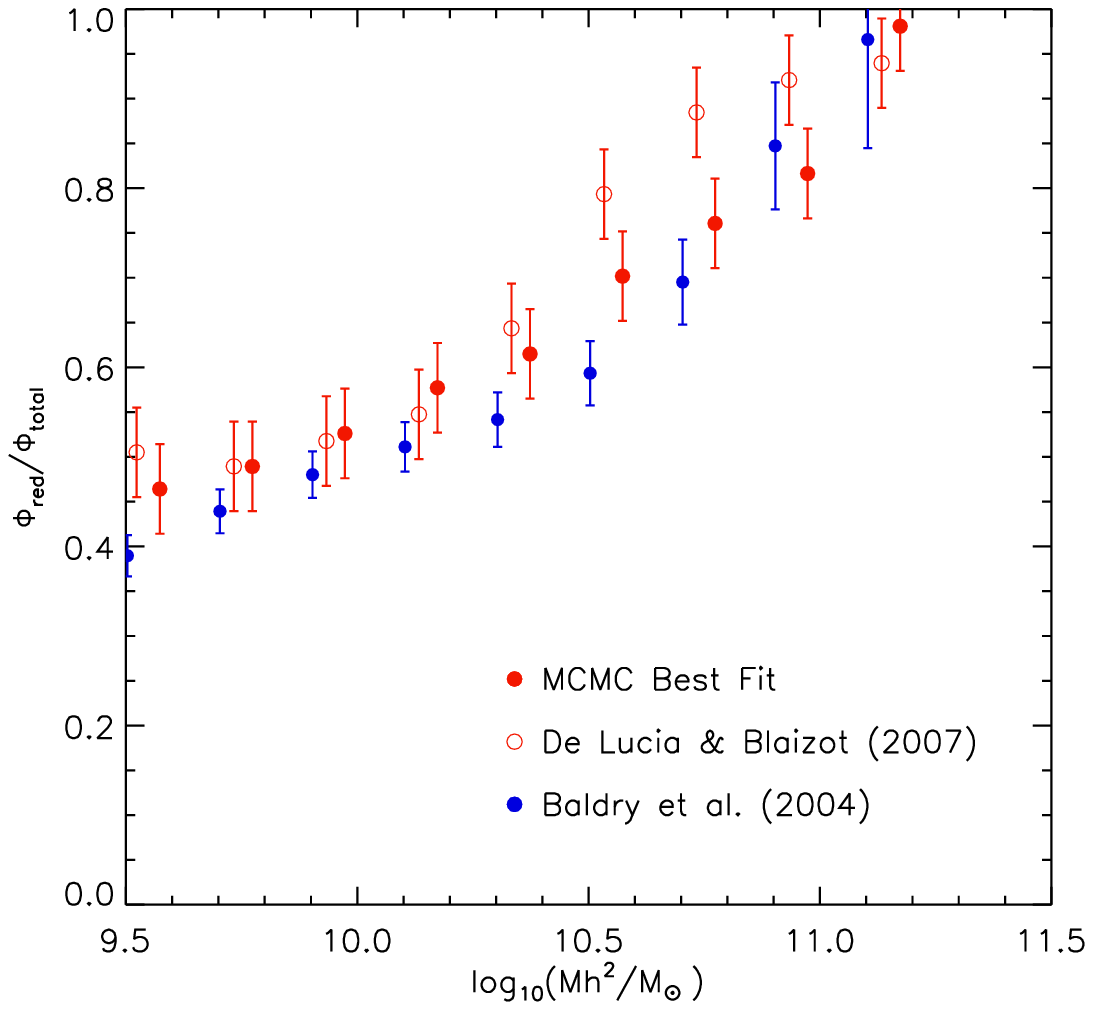}
\caption{The top panel shows the $B$-$V$ colour-stellar mass relation
  for the galaxies in our best fit model.  The solid line represents
  the division between the red and blue populations in
  \citet{Weinmann2006a}. The predicted fraction of red galaxies as a 
  function of stellar mass is showed in the bottom panel. The original
  DLB07 model (open red circles) is compared with our best fit model
  (filled red circles) and observational data from \citet{Baldry2004}
  (filled blue squares).}
\label{fig:bestfitcolorfract}
\end{figure}

In Fig. \ref{fig:bestfitcolorfract} we show the predicted galaxy
colours in our best fit model, the $B$-$V$ colour-stellar mass
relation on the top panel and the fraction of red over the total number of
galaxies as a function of stellar mass in the bottom panel.  Our best
fit correctly reproduces the fraction of red galaxies by slightly
increasing the number of blue galaxies around $L_{\star}$ compared to
DLB07.

The colour-stellar mass relation also shows improvements,
keeping the bimodality between the red and the blue galaxies, but
increasing the slope of each population as suggested by observations.
Nevertheless, near the lower-mass limit we impose in our study a
population of red dwarfs starts to emerge, representing the highest
number density peak in the red population.  This is in disagreement
with observations, where the majority of the red galaxies are massive,
and the dwarfs are predominately blue.  We address this problem, and
explore possible solutions in section \ref{conclusions}.

\subsubsection{The Black Hole-Bulge Mass Relation}
\label{bestfitbhbm}

The best-fit BH-bulge mass relation is almost unchanged from that
shown in Figure~\ref{fig:originalbbh} and so we do not repeat it here.
There is enough freedom in the model to allow the AGN parameters to
adjust themselves to recover the correct BH masses, despite the
differences in the SN parameters between DLB07 and out best fit.


\subsection{The Galaxy Stellar Mass Function}
\label{bestfitsmf}

\begin{figure}
\centering
\includegraphics[width=8.4cm]{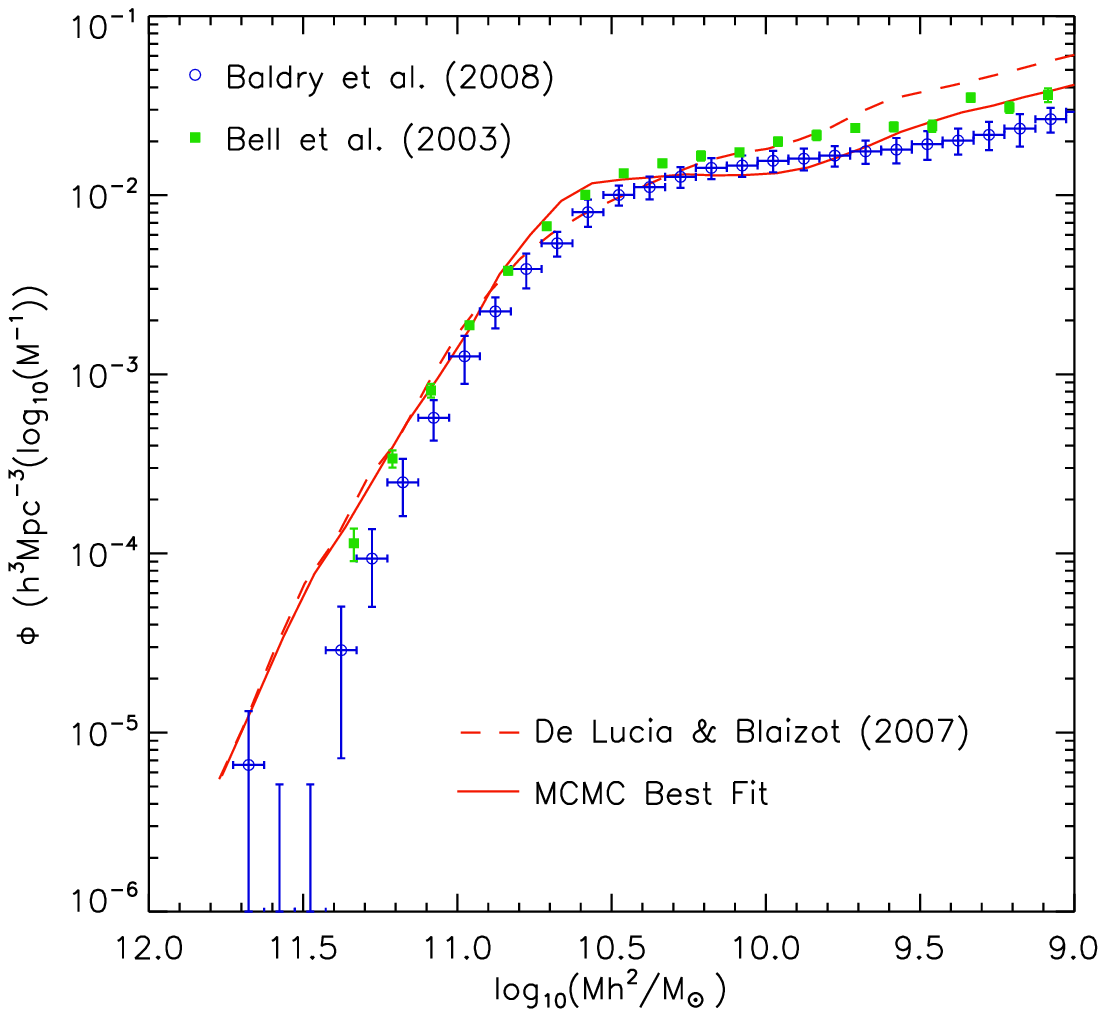}
\caption{Comparison for the predicted stellar mass function at z=0
  from DLB07 (dashed red line) and our best-fit model (solid red line)
  with observations from \citet{Baldry2008} (blue open circles) and
  \citet{Bell2003}(green filled squares).}
\label{fig:bestfitsmf}
\end{figure}

As discussed in section \ref{kband}, the stellar mass function is one
of the most fundamental properties of a galaxy population, but it is
difficult to derive accurately from observations.  In
Fig.~\ref{fig:bestfitsmf} we show how our best-fit model and DLB07
masses compare with observationally-derived stellar mass functions
from \citet{Bell2003} and \citet{Baldry2008}.  The latter is one of
the most robust mass derivations, using the New York University
Value-Added Galaxy Catalogue that combines 4 different methods for
determining galaxy masses from SDSS data.  The error bars in the
figure span the maximum and minimum mass estimates from that analysis.

After all the data sets are converted into the same IMF \citep[that
  of][]{Chabrier2003}, the comparison between our best-fit model and
the masses derived from \citet{Bell2003} in Fig.~\ref{fig:bestfitsmf}
shows the same behaviour as the $K$-band luminosity function.  With
our more effective supernova feedback the excess of dwarf galaxies
largely disappears, but there is a slight excess of $L\star$ galaxies.
Small differences might arise when comparing both the $k$-band and
stellar mass functions from the model and a specific observational
data set, even though similar stellar population synthesis models were
used.  This is because, to convert stellar mass into luminosity, one
should have knowledge about the age and metallicity of the galaxy
stellar population.  These quantities are directly available in the
model, but in observations they are difficult to derive and are
subject to large uncertainties.

Also shown in Fig.~\ref{fig:bestfitsmf} is a comparison of the
predicted masses with data from \citet{Baldry2008}.  A systematic
difference between the model and the data is evident, with the former
predicting a much larger number of galaxies on and above $L\star$.
While the horizontal error bars plotted in Fig.~\ref{fig:bestfitsmf}
for \citet{Baldry2008} reflect only the bin size, the authors refer to
differences as large as 0.15 dex in mass estimates from the different
methods.  Another recent work \citep{Conroy2008} points to even larger
errors of up to 0.3 dex that may result from imprecise modelling of
key phases of stellar evolution.  If these uncertainties lead to the
observationally-derived masses of large galaxies being underestimated
by about 0.15 dex, then our best fit and observations from
\citet{Baldry2008} would be in extremely good agreement throughout the
whole mass range.

The differences between data sets highlight the need for caution when
galaxy formation models are compared with observations.  In principle
one should expect the properties and allowed parameter ranges to
change if either the stellar population synthesis or the dust model
need to be readjusted.  In this paper we have chosen to fix these so as to
focus our study on the parameters controlling the most basic properties
of the semi-analytic model: star formation and feedback.

\section{Discussion and Conclusions}
\label{conclusions}

Since they were introduced as a new technique to understand galaxy
formation \citep{White1978,White1991,Lacey1991,Cole1991},
semi-analytic models have always lacked a proper statistical analysis of
the allowed range of their free parameters and a consistent way to
test the goodness of the fits produced.

To overcome this weakness we have implemented a Monte Carlo Markov
Chain parameter estimation technique into the \citet{Delucia2007}
semi-analytic model, to obtain the best values and confidence limits
for the 6 free parameters in the model responsible for shaping the
stellar mass function and for the colours of galaxies.  Comparing the
model with three different observational constraints separately: the
combined $K$-band luminosity function from
\citet{Cole2001,Bell2003,Jones2006}, galaxy colours from
\citet{Baldry2004} and the BH-bulge mass relation from
\citet{Haring2004}, we are able to identify which particular
parameters (and hence which galaxy formation processes) are
responsible for each individual property and which show correlations
and degeneracies.

Combining the three observational tests, we are able to fully
constrain the model parameters, obtaining a best fit and confidence
limits within the very limited region of acceptable likelihood.  Our
best model is given by: $\alpha_{SF}=0.039_{-0.019}^{+0.091}$,
$k_{AGN}=(5.0_{-2.6}^{+6.0})\times10^{-6}$,
$f_{BH}=0.032_{-0.018}^{+0.083}$,
$\epsilon_{disk}=10.28_{-5.85}^{+14.09}$,
$\epsilon_{halo}=0.53_{-0.27}^{+0.64}$ and
$\gamma_{ej}=0.42_{-0.34}^{+0.37}$).  As shown in table
\ref{table:margestats}, all the parameters in the original model,
except the supernova reheating efficiency fall within our 2-$\sigma$
confidence limits. Our best fit maintains the values for the star
formation efficiency and for the AGN quasar mode parameters, while
increasing the SN gas reheating and ejection and decreasing the AGN
radio mode and gas reincorporation efficiencies.  For our preferred
set of parameters the model has a likelihood of
$\pi(x_i)={L}_{\rm(Mass)} \times {L}_{\rm(Colour)} \times
{L}_{(\mathrm{BH-Bulge})}=0.037$.  This value means that the best-fit
solution is incompatible with the three combined observations at the
2-$\sigma$ level.

In this paper, we have used the Millennium Simulation which
  adopts a $\Lambda$CDM cosmology.  It is possible that the Universe
  may be better described by an alternative cosmology with fewer
  low-mass halos.  However, our purpose here is to
  try to find an astrophysical solution that is compatible with
  $\Lambda$CDM.

As discussed in previous chapters, observational uncertainties,
principally that associated with stellar population synthesis
and dust modelling, make it premature to conclude that the
DLB07 formalism is ruled out: shifting the observed
black-hole/bulge mass ratio by 0.15 dex raises the best-fit
likelihood to 0.07, which is marginally acceptable.  Nevertheless,
the apparent incompatibility between the black hole-bulge mass
relation and the other constraints indicates that the black hole
growth treatment in the model might be too simplistic, in particular
by assuming that there is no feedback from the quasar mode, when it
seems to be required to reproduce the X-ray luminosity
  function of halos \citep{Bower2008,Short2008}.  As discussed below,
some additional recipes might also need to be included for the model
to better reproduce observational luminosities and colours, which
could in principle increase the likelihood of the best fit model.

We produce a $K$-band luminosity function for our best parameter
values that improves the agreement with observations at the
low-luminosity end.  This is achieved by taking a higher heating
efficiency from SN and a lower reincorporation rate of gas ejected
from the halo.  This reduces the amount of cold gas available to form
stars, avoiding the excess of faint galaxies in the original recipe.
More effective SN feedback has been proposed in the past.
For example, \citet{Bertone2007} studied a wind model that improved the
number density of dwarfs for both the mass and luminosity function 
while improving the distribution of metals. 
However, the high value of ejection that is required by our model seems to be 
unrealistic when compared with observations \citep{Martin1999}.
This indicates that additional processes such as disruption of
 satellites through tidal effects might need to be included
\citep{Bullock2001,Taylor2001,Benson2002a,Monaco2006,
  Weinmann2006b,Murante2007,Henriques2008,Somerville2008}.

For both the luminosity-function and colour constraints, the SN
reheating and ejection parameters are strongly correlated, which we
interpret as an upper virial velocity limit of 140\,km\,s$^{-1}$ for
galaxies that can eject mass via SN heating. Since we need to assume
a stronger SN feedback to decrease the faint end of the $K$-band
luminosity function, this relatively low value assures that our SN
feedback stops being effective for galaxies with masses above 
$M_{\star}\approx 10^{10.5}M_{\odot}$.  This is the only way to
ensure that enough stars will form in brighter galaxies, and also
produces more blue galaxies around $L_{\star}$ than the original DLB07
parameters. 

Significant correlations exist between the three parameters governing
SN feedback , $\epsilon_{\rm disk}$, $\epsilon_{\rm halo}$ \&
$\gamma_{\rm ej}$, even in the combined analysis.  This suggests that
the model could be rewritten with one or two fewer free
parameters. However this degeneracy could in principle be broken if
the metallicity of gas and stars were to be considered.

The model with the original DLB07 parameters correctly predicts the
bimodality in the colour-stellar mass relation, however, it has
difficulties in matching the exact number of the blue and red sequence
galaxies. Our best-fit model correctly predicts the relative fraction of
galaxies in each colour population, however the early cutoff on SN
feedback leads to an excess of galaxies with masses between $10^{10.5}$ and
$10^{11.0}$ \Msun. Furthermore, as the original model, 
it shows a large population of small red galaxies 
in the $B-V$ colour-stellar mass plot in contradiction 
with observations.  

The problems with low-mass galaxy colours in semi-analytic models have
been identified in the past, particularly the excess of red dwarfs
\citep{Croton2006,Baldry2006}.  Possible solutions might include the
delayed stripping of gas from satellites after their dark matter halo
is disrupted (allowing them to cool gas, form stars and stay blue for
longer \citealt{Font2008}). Or again, tidal disruption of dwarfs, which
would affect mostly red, satellite galaxies. This would move them
to even lower masses, below $10^{9.0}M_{\odot}$ (where an upturn
in the stellar mass function is seen) and produce intra-cluster light
\citep{Weinmann2006b,Henriques2008, Somerville2008}.

The purpose of this paper is to show that MCMC parameter estimation
techniques, adapted from those used in cosmology, can be used to map
out likelihood contours in the parameter space of semi-analytic models
of galaxy formation.  For this particular analysis we chose the
formalism of \citet{Delucia2007}, but the same method could equally be
applied to other models.

In the future, we would like to extend the method to undertake model
selection, providing an objective measure of the relative value of
models with different numbers of free parameters.

\section*{Acknowledgements}
We thank all the members of the Sussex Survey Science Centre which
joint expertise helped developing the innovative idea in this paper,
specially David Parkinson, untiring on spreading his great Bayesian
knowledge.  We are grateful to Volker Springel and Gabriella De Lucia
for providing us with the Munich semi-analytic code and for supporting
our use of it.  We thank also Ivan Baldry for providing us the data on
galaxy colours and masses and for his always helpful comments.

We would like to thank the anonymous referee for comments that helped
to clarify some of the discussion in the paper.

The computations developed for this work, were performed in the Virgo
Consortium cluster of computers, COSMA. The authors would like to
thank Lydia Heck for its great technical knowledge about COSMA and
constant feedback without which this work could not have been done. 

BH acknowledges the support of his PhD scholarship from the Portuguese
Science and Technology Foundation, truly thankful to this institution
and all the tax payers in Portugal, for keeping the science dream
alive in troubled times.

\bibliographystyle{mn2e}
\bibliography{references}

\label{lastpage}

\end{document}